\documentclass[12pt]{article}
\usepackage{amsmath}
\usepackage{amssymb}

\DeclareMathOperator{\csch}{csch}
\DeclareMathOperator{\sech}{sech}
\newcommand{\half}{\frac{1}{2}}
\newcommand{\ah}{\hat{A}}
\newcommand{\ph}{\hat{P}}
\newcommand{\htilde}{\tilde{H}}
\newcommand{\vt}{\tilde{V}}
\newcommand{\wt}{\tilde{W}}
\newcommand{\et}{\tilde{E}}
\newcommand{\bh}{\hat{B}}
\newcommand{\phit}{\tilde{\phi}}
\newcommand{\ca}{{\cal A}}
\newcommand{\cb}{\bar{c}}
\newcommand{\ri}{{\rm i}}

\title{
Isospectrality of conventional and new extended potentials, second-order supersymmetry and role of $\cal PT$ symmetry}
\author{B.\ Bagchi $^a$, C.\ Quesne $^{b,}$\thanks{Corresponding author.
{\sl E-mail addresses}: bbagchi123@rediffmail.com (B.\ Bagchi), cquesne@ulb.ac.be (C.\
Quesne), rajdaju@rediffmail.com (R.\ Roychoudhury).}\ , R.\ Roychoudhury $^c$\\ 
{\small\sl $^a$ Department of Applied Mathematics, University of Calcutta,} \\ 
{\small\sl 92 Acharya Prafulla Chandra Road, Kolkata 700 009, India}\\ 
{\small\sl $^b$ Physique Nucl\'eaire Th\'eorique et Physique Math\'ematique,  Universit\'e Libre de Bruxelles,} \\ 
{\small\sl Campus de la Plaine CP229, Boulevard~du Triomphe, B-1050
Brussels, Belgium}\\
{\small\sl $^c$ Physics and Applied Mathematics Unit, Indian Statistical Institute, Kolkata
700 035, India}}
\date{ }
\begin{document}
\baselineskip=22pt plus 1pt minus 1pt
%%%%%%%%%%%%%%%%%%%%%%%%%%%%%%%%%%%%%%%%%%%%%%%%%%%%%%%%%%
\maketitle

\begin{abstract} 
We develop a systematic approach to construct novel completely solvable rational potentials. Second-order supersymmetric quantum mechanics dictates the latter to be isospectral to some well-studied quantum systems. $\cal PT$ symmetry may facilitate reconciling our approach to the requirement that the rationally-extended potentials be singularity free. Some examples are shown.
\end{abstract}



%
%========================================================================
%
\newpage
\section{Introduction}

The search for exactly solvable potentials has been a subject of intense research since the advent of quantum mechanics because they may provide a conceptual understanding of some physical phenomena, as well as a testing ground for some approximation schemes. This has continued to be so after the generalization of quantum theory for the case of complex $\cal PT$-symmetric \cite{bender} or pseudo-Hermitian \cite{mosta} Hamiltonians.\par
%
%-------------------------------------------------------------------------------------------------------
%
Over the years supersymmetric quantum mechanics (SUSYQM) has emerged as one of the most insightful tools towards construction of Hamiltonians with a prescribed spectrum starting from some given exactly solvable form \cite{cooper, junker96, bagchi00a, mielnik04}. The key technique such as the factorization method (sometimes couched in the language of intertwining relationships amongst operators inducing factorization) has enabled one to uncover many useful properties of quantum mechanics. SUSY systems are appropriately described by a superpotential descendant of the given Schr\"odinger potential, which in turn furnishes the so-called SUSY partner Hamiltonians. In one-dimensional systems, strict isospectrality may be present in the first-order systems (broken SUSY) when the factorization energy $E$ is smaller than the ground-state energy of the starting Hamiltonian \cite{mielnik84, sukumar}. In such a context, the potential and eigenfunctions of the partner Hamiltonian are known in terms of analytic expressions, which, however, may be a good deal more complicated than the corresponding ones of the initial Hamiltonian \cite{contreras}. Interestingly one can look for extensions to higher-order SUSY theories by resorting to higher-derivative versions of the factorization operators. These offer the scope of relating to non-trivial quantum systems as has been found in a number of recent works \cite{andrianov, samsonov, bagchi99, fernandez}.\par
%
%-------------------------------------------------------------------------------------------------
%
Against this background, the recent introduction of two new classes of exceptional orthogonal polynomials \cite{gomez08} and their occurrence in the bound-state wavefunctions of two novel rational potentials isospectral to some well-known conventional ones \cite{cq} have led us to re-examine the construction of such pairs of partner potentials. These examples have suggested us an alternative approach to the usual one, which consists in searching for a reduction of the initial Schr\"odinger equation general solution to some elementary function \cite{gomez04}.\par
%
%----------------------------------------------------------------------------------------------------------------
%
Taking advantage of such an experience, we plan to enquire here how flexible is the solution of the superpotential that forms a given SUSY quantum model. Employing the factorization method, which was generalized by Gendenshtein \cite{gendenshtein} in terms of the reparametrization of the potential ensuring thereby an algebraic construction of the latter, we will generate new pairs of SUSY Hamiltonians by effecting a reparametrization of the starting Schr\"odinger potential. We will show how the family of such pairs can be substantially enlarged should we redefine the existing superpotential in terms of new couplings and allow for the presence of some rational function in it. The latter plays an interesting dual role: it not only allows coincidence of one of the new partner Hamiltonians with the starting one up to some reparametrization of couplings, but also forces the other partner to emerge as a superposition of the initial exactly solvable potential (with its parameters left undisturbed) and some additional rational terms, thus representing an extended family of exactly solvable rational potentials. Such an extended potential has its spectrum unaltered vis-\`a-vis the exactly solvable one from which it is obtained, while the corresponding bound-state wavefunctions turn out to be expressible in terms of rather simple polynomials. It is important to stress that a second-order supersymmetric (SSUSY) transformation lies at the very root of the observed isospectrality. Furthermore, our findings show that for some potentials $\cal PT$ symmetry \cite{bender} enters the picture in a natural way to render the derived potential singularity free.\par
%
%================================================================
% 
\section{Generalized P\"oschl-Teller potential}
We begin with the concrete example of the generalized P\"oschl-Teller (GPT) potential as our first model to address the issue of constructing an extended class of rational potentials.\par
%
%+++++++++++++++++++++++++++++++++++++++++++++++++++++++++
%
\subsection{Extended family of rational potentials}
The GPT potential reads in standard notations \cite{cooper}
\begin{equation}
  V_{A,B}(x) = [B^2 + A(A+1)] \csch^2 x - B (2A+1) \csch x \coth x  \label{eq:GPT-V}
\end{equation}
and is defined on the half-line $0 < x < \infty$. It is repulsive at the origin for $B > A+1 > 1$ but supports a finite number of bound states, whose energies are given by
\begin{equation}
  E^{(A)}_{\nu} = - (A - \nu)^2, \qquad \nu = 0, 1, \ldots, \nu_{\rm max} \quad (A - 1 \le \nu_{\rm max} < A).
  \label{eq:GPT-E}
\end{equation}
The associated wavefunctions, vanishing at the origin and decaying exponentially at infinity, can be expressed in terms of Jacobi polynomials as
\begin{equation}
  \psi_{\nu}^{(A,B)}(x) = {\cal N}^{(A,B)}_{\nu} (\cosh x - 1)^{\half(B-A)} (\cosh x + 1)^{-\half(B+A)} 
  P^{\left(B-A-\half, -B-A-\half\right)}_{\nu}(\cosh x),  \label{eq:GPT-psi} 
\end{equation}
where
\begin{equation*}
  {\cal N}^{(A,B)}_{\nu} = 2^A \left(\frac{\nu! (2A-2\nu) \Gamma\left(B+A-\nu+\half\right)}
  {\Gamma\left(B-A+\nu+\half\right) \Gamma(2A-\nu+1)}\right)^{1/2}  
\end{equation*}
denotes a normalization constant\footnote{Such a picture is also valid whenever the parameters vary in the range $A+1 > B > A + \half > \half$ or $A+\half > B > A > 0$, where the potential is weakly attractive at the origin. Among the two square integrable solutions of the Schr\"odinger equation corresponding to a given $E^{(A)}_{\nu}$, it is customary to consider the function (\ref{eq:GPT-psi}) as the physical one because it is the most regular.}.\par
%
%----------------------------------------------------------------------------------------------------------------------------
%
In the standard SUSYQM approach to such a potential, which incidentally is also shape invariant, deletion of the ground-state energy $E^{(A)}_0$ yields a partner potential $V_{A-1,B}(x)$ of the same kind by using a superpotential having the form $A \coth x - B \csch x$ \cite{cooper}.\par
%
%----------------------------------------------------------------------------------------------------------
%
Motivated by the previous results on the non-uniqueness of factorization schemes which also have relevance to the construction of rational potentials \cite{mielnik84, gomez04, bargmann, mitra, junker97}, let us now ask the following question: can we modify the above superpotential for the GPT potential in such a way that the starting potential $V^{(+)}(x) = W^2 - W' + E$ ($E$ denoting the factorization energy) continues to belong to the GPT family up to some reparametrization of coefficients, while its partner $V^{(-)}(x) = W^2 + W' + E$ differs from $V_{A,B}(x)$ in the presence of some additional rational terms? Taking cue from the observation that $z = \cosh x$ is the basic variable appearing in the Jacobi polynomial that controls (\ref{eq:GPT-psi}), we propose for such a superpotential the following form
\begin{equation}
  W(x) = a \coth x + b \csch x - \frac{\sinh x}{\cosh x + c},  \label{eq:W}
\end{equation}
where the constants $a$, $b$, $c$ are to be determined from suitable inputs.\par
%
%---------------------------------------------------------------------------------------------------------
%
Straightforward calculations using (\ref{eq:W}) lead to
\begin{equation}
  V^{(+)}(x) = [a(a+1) + b^2] \csch^2 x + (2a+1) b \csch x \coth x + \frac{(2a-1) c - 2b}{\cosh x + c} +
  (a-1)^2 + E  \label{eq:V+}
\end{equation}
and
\begin{equation}
\begin{split}
  V^{(-)}(x) &= [a(a-1) + b^2] \csch^2 x + (2a-1) b \csch x \coth x + \frac{(2a-3) c - 2b}{\cosh x + c} \\  
  &\quad + \frac{2(c^2 - 1)}{(\cosh x + c)^2} + (a-1)^2 + E.
\end{split}  \label{eq:V-}
\end{equation}
Elimination of the term proportional to $(\cosh x + c)^{-1}$ from $V^{(+)}(x)$ can be achieved provided we choose $c = 2b/(2a-1)$. Furthermore, the first two terms of $V^{(-)}(x)$ coincide with $V_{A,B}(x)$, given in (\ref{eq:GPT-V}), if the two conditions
\begin{equation}
  a(a-1) + b^2 = B^2 + A(A+1), \qquad (2a-1) b = - B (2A+1)  \label{eq:ab}
\end{equation}
are fulfilled. Finally, the vanishing of the constants terms in (\ref{eq:V+}) and (\ref{eq:V-}) imposes that the factorization energy be given by $E = - (a-1)^2$.\par
%
%-----------------------------------------------------------------------------------------------------------
%
The two restrictions in (\ref{eq:ab}) lead to four possible solutions for $a$ and $b$. The first two are given by $a = \pm \left(B \pm \half\right)$, $b = \mp \left(A + \half\right)$, where we take all upper or all lower signs, while the two remaining ones can be obtained from them by permuting $A + \half$ with $B$. As a result, $c$ becomes $c = - (2A+1)/(2B)$ or $c = - 2B/(2A+1)$, respectively. Only for the former $c$ value can we assert that the extended potential $V^{(-)}(x)$ will be free from poles on the half-line. We therefore restrict ourselves to such a value and arrive at the following results:
\begin{align}
  V^{(+)}(x) &= V_{A, B\pm1}(x) = [(B\pm1)^2 + A(A+1)] \csch^2 x - (B\pm1) (2A+1) \csch x \coth x,
        \label{eq:V+bis} \\
  V^{(-)}(x) &= V_{A, B, {\rm ext}}(x) = V_{A, B}(x) + \frac{2(2A+1)}{2B \cosh x - 2A-1} - \frac{2[4B^2 -
        (2A+1)^2]}{(2B \cosh x - 2A-1)^2},  \label{eq:V-bis} 
\end{align}
with
\begin{equation}
\begin{split}
  W(x) &= \pm \left(B \pm \half\right) \coth x \mp \left(A + \half\right) \csch x - \frac{2B \sinh x}
        {2B \cosh x - 2A-1}, \\
  E &= - \left(B \mp \half\right)^2.  
\end{split}  \label{eq:W-E}
\end{equation}
We have therefore found two distinct solutions to the above-mentioned question: going from a conventional GPT potential to an extended potential $V_{A, B, {\rm ext}}(x)$ can be achieved in a first-order SUSYQM setting provided we start either from $V_{A, B+1}(x)$ or from $V_{A, B-1}(x)$ (but in no way if we start from a potential $V_{A, B}(x)$ with the same parameters). It is also worthwhile to note that although the GPT potential $V_{A, B}(x)$ is not defined for $B = A + \half$, the same is not true for $V_{A, B, {\rm ext}}(x)$, which can easily be shown to be equivalent to the well-behaved  conventional GPT potential $V_{A-1, A+\frac{3}{2}}(x)$, with a correspondence between their respective bound-state energies and wavefunctions.\par
%
%------------------------------------------------------------------------------------------------------------------
%
The facts that the GPT energy eigenvalues (\ref{eq:GPT-E}) are independent of $B$ (provided this parameter satisfies the condition $B > A+1$) and that the factorization energy $E$, given in (\ref{eq:W-E}), is smaller than $E^{(A)}_0$ make it plausible that the partner potential $V^{(-)}(x)$ in (\ref{eq:V-bis}) be strictly isospectral to any choice for $V^{(+)}(x)$ in (\ref{eq:V+bis}) (as well as to $V_{A, B}(x)$ in (\ref{eq:GPT-V})). This is actually confirmed by determining the factorization function $\phi(x)$, which is such that $W = - \phi'/\phi$. The result reads
\begin{equation*}
  \phi(x) \propto (\cosh x - 1)^{\mp\half\left(B-A\pm\half-\half\right)} 
  (\cosh x + 1)^{\mp\half\left(B+A\pm\half+\half\right)} (2B \cosh x - 2A - 1),
\end{equation*}
where the upper (resp.\ lower) signs correspond to $V_{A, B+1}(x)$ (resp.\ $V_{A, B-1}(x)$) in (\ref{eq:V+bis}). We note that for the former choice, $\phi(x)$ decreases exponentially for $x \to \infty$, but increases as $x^{-(B-A)}$ for $x \to 0$, while for the latter choice it vanishes as $x^{B-A-1}$ for $x \to 0$, but increases exponentially for $x \to \infty$, thus showing in both cases that neither $\phi$ nor $\phi^{-1}$ is normalizable, as is required by the general theory \cite{sukumar}.\par
%
%++++++++++++++++++++++++++++++++++++++++++++++++++++++++++++++
% 
\subsection{Determination of bound-state wavefunctions}

The bound-state wavefunctions $\psi^{(-)}_{\nu}(x) \equiv \psi^{(A, B, {\rm ext})}_{\nu}(x)$ of the extended GPT potential can be determined by using the standard intertwining relations $H^{(+)} \ah^{\dagger} = \ah^{\dagger} H^{(-)}$, $\ah H^{(+)} = H^{(-)} \ah$ \cite{cooper, junker96, bagchi00a, mielnik04}, satisfied by the partner Hamiltonians
\begin{equation*}
  H^{(+)} = \ah^{\dagger} \ah = - \frac{d^2}{dx^2} + V^{(+)}(x) - E, \qquad H^{(-)} = \ah \ah^{\dagger} = 
  - \frac{d^2}{dx^2} + V^{(-)}(x) - E. 
\end{equation*}
Here $\ah^{\dagger} = - \frac{d}{dx} + W(x)$ and $\ah = \frac{d}{dx} + W(x)$. Hence choosing, for instance, $V_{A, B+1}(x)$ in (\ref{eq:V+bis}), we can write
\begin{equation}
  \psi^{(-)}_{\nu}(x) = \frac{1}{\sqrt{\varepsilon_{\nu}}} \ah \psi^{(+)}_{\nu}(x), \qquad \psi^{(+)}_{\nu}(x) = 
  \psi^{(A, B+1)}_{\nu}(x), \qquad \varepsilon_{\nu} = E^{(A)}_{\nu} - E,  \label{eq:psi-} 
\end{equation}
for any $\nu = 0$, 1,~\ldots, $\nu_{\rm max}$.\par
%
%------------------------------------------------------------------------------------------------------------
%
On using Eq.\ (\ref{eq:W-E}), as well as Eq.\ (\ref{eq:GPT-psi}) with $B+1$ substituted for $B$, we can rewrite (\ref{eq:psi-}) as
\begin{equation}
  \psi^{(-)}_{\nu}(x) = \frac{{\cal N}^{(+)}_{\nu}}{\sqrt{\varepsilon_{\nu}}} 
  \frac{(z-1)^{\half\left(\alpha+\half\right)} (z+1)^{\half\left(\beta+\half\right)}}{\beta+\alpha 
  - (\beta-\alpha)z} {\cal O}^{(\alpha,\beta)}_z P^{(\alpha+1, \beta-1)}_{\nu}(z),  \label{eq:psi-bis}
\end{equation}
where we have defined ${\cal N}^{(+)}_{\nu} = {\cal N}^{(A,B+1)}_{\nu}$, $z = z(x) = \cosh x$, $\alpha = B - A - \half$, $\beta = - B - A - \half$ and
\begin{equation*}
  {\cal O}^{(\alpha,\beta)}_z = [\beta+\alpha - (\beta-\alpha)z] \left((z-1) \frac{d}{dz} + \alpha + 1\right)
  + (\beta-\alpha) (z-1).
\end{equation*}
\par
%
%-----------------------------------------------------------------------------------------------------------
%
The action of the first-order differential operator ${\cal O}^{(\alpha,\beta)}_z$ on the Jacobi polynomial $P^{(\alpha+1, \beta-1)}_{\nu}(z)$ can be inferred from some known differential and recursion relations for the latter \cite{abramowitz}. The result can be written as
\begin{equation}
  {\cal O}^{(\alpha,\beta)}_z P^{(\alpha+1, \beta-1)}_{\nu}(z) = - 2 (\alpha-\beta) (\nu+\alpha)
  \ph^{(\alpha, \beta)}_{\nu+1}(z),  \label{eq:O}
\end{equation}
where $\ph^{(\alpha, \beta)}_{\nu+1}(z)$ is a $(\nu+1)$th-degree polynomial, defined by
\begin{equation}
  \ph^{(\alpha, \beta)}_{\nu+1}(z) = - \half (z-b)  P^{(\alpha, \beta)}_{\nu}(z) + 
  \frac{b P^{(\alpha, \beta)}_{\nu}(z) - P^{(\alpha, \beta)}_{\nu-1}(z)}{\alpha + \beta + 2\nu}, \qquad
  b \equiv \frac{\beta+\alpha}{\beta-\alpha},  \label{eq:definition} 
\end{equation}
in terms of some Jacobi ones.\par
%
%------------------------------------------------------------------------------------------------------------------
%
At this stage, SUSYQM can be profitably used to construct orthonormal bound-state wavefunctions $\psi^{(-)}_{\nu}(x)$ in terms of $\ph^{(\alpha, \beta)}_{\nu+1}(z)$. Inserting (\ref{eq:O}) in (\ref{eq:psi-bis}) and taking all previous definitions into account indeed leads to
\begin{equation}
  \psi^{(-)}_{\nu}(x) = {\cal N}^{(-)}_{\nu} \frac{(\cosh x - 1)^{\half(B-A)} (\cosh x + 1)^{-\half(B+A)}}
  {2B \cosh x - 2A - 1} \ph^{\left(B-A-\half, -B-A-\half\right)}_{\nu+1}(\cosh x),  \label{eq:ext-psi}
\end{equation}
where
\begin{equation*}
\begin{split}
  {\cal N}^{(-)}_{\nu} &= - 4B \left(\frac{B-A+\nu-\half}{B+A-\nu-\half}\right)^{1/2} {\cal N}^{(+)}_{\nu} \\
  &= - 2^{A+2} B \left(\frac{\nu!\, (2A-2\nu) \left(B+A-\nu+\half\right) \Gamma\left(B+A-\nu-\half\right)}
        {\left(B-A+\nu+\half\right) \Gamma\left(B-A+\nu-\half\right) \Gamma\left(2A-\nu+1\right)}\right)
        ^{1/2}.  
\end{split}
\end{equation*}
\par
%
%++++++++++++++++++++++++++++++++++++++++++++++++++++++++
%
\subsection{Isospectrality and SSUSY}

Let us now focus on the isospectrality issue of the conventional and extended potentials with the same parameters $A$, $B$. Since we have been able to go from $V_{A, B\pm1}(x)$ to $V_{A,B,{\rm ext}}(x)$ by a first-order SUSYQM transformation, we may try to convert $V_{A,B}(x)$ into $V_{A,B\pm1}(x)$ in a preliminary step. As is easily seen, such a transformation can be carried out in a first-order SUSYQM setting by assuming
\begin{equation}
\begin{split}
  &\htilde^{(+)} = \bh^{\dagger} \bh = - \frac{d^2}{dx^2} + \vt^{(+)}(x) - \et, \qquad \htilde^{(-)} 
          = \bh \bh^{\dagger} = - \frac{d^2}{dx^2} + \vt^{(-)}(x) - \et,  \\
  &\vt^{(+)}(x) = V_{A,B}(x) = \wt^2 - \wt' + \et, \qquad \vt^{(-)}(x) = V_{A,B\pm1}(x) = \wt^2 + \wt' + 
          \et,  \\
  &\bh^{\dagger} = - \frac{d}{dx} + \wt(x), \qquad \bh = \frac{d}{dx} + \wt(x), 
\end{split}  \label{eq:Vtilde}
\end{equation}
with a superpotential and a factorization energy given by
\begin{equation*}
  \wt(x) = \mp \left(B\pm\tfrac{1}{2}\right) \coth x \pm \left(A+\tfrac{1}{2}\right) \csch x \qquad \text{and} 
  \qquad \et = - \left(B\pm\tfrac{1}{2}\right)^2,
\end{equation*}
respectively. As before, $\et < E^{(A)}_0$ and the corresponding factorization function
\begin{equation*}
  \phit(x) \propto (\cosh x - 1)^{\pm\half\left(B-A\pm\half-\half\right)}
  (\cosh x + 1)^{\pm\half\left(B+A\pm\half+\half\right)}
\end{equation*}
is such that neither $\phit$ nor $\phit^{-1}$ is normalizable, thereby ensuring the (strict) isospectrality of this step too.\par
%
%-------------------------------------------------------------------------------------------------
%
Since from (\ref{eq:V+bis}) and (\ref{eq:Vtilde}), it follows that $\vt^{(-)}(x) = V^{(+)}(x)$, the two first-order SUSY systems $\left(\htilde^{(+)}, \htilde^{(-)}\right)$ and $\left(H^{(+)}, H^{(-)}\right)$ can be glued together so as to get a reducible SSUSY one $\left(h^{(1)}, h^{(2)}\right)$ \cite{andrianov, samsonov, bagchi99, fernandez}. In such a framework, the two Hamiltonians $h^{(1)}$ and $h^{(2)}$ with corresponding potentials $V^{(1)}(x)$ and $V^{(2)}(x)$, respectively, intertwine with some second-order differential operators
\begin{equation}
  \ca^{\dagger} = \frac{d^2}{dx^2} - 2p(x) \frac{d}{dx} + q(x), \qquad \ca = \frac{d^2}{dx^2} + 2p(x) 
  \frac{d}{dx} + 2p'(x) + q(x),  \label{eq:intertwine}
\end{equation}
as $\ca h^{(1)} = h^{(2)} \ca$ and $\ca^{\dagger} h^{(2)} = h^{(1)} \ca^{\dagger}$, so that the functions $p(x)$, $q(x)$ and the potentials $V^{(1,2)}(x)$ are constrained by the relations
\begin{equation*}
\begin{split}
  & q(x) = - p' + p^2 - \frac{p''}{2p} + \left(\frac{p'}{2p}\right)^2 - \frac{\cb^2}{16p^2},  \\
  & V^{(1,2)}(x) = \mp 2p' + p^2 + \frac{p''}{2p} - \left(\frac{p'}{2p}\right)^2 + \frac{\cb^2}{16p^2}, 
\end{split}
\end{equation*}
where $\cb$ is some integration constant.\par
%
%---------------------------------------------------------------------------------------------------------------
%
The relation between both approaches follows from the equations
\begin{equation}
  h^{(1)} = \htilde^{(+)} + \frac{\cb}{2}, \qquad h^{(2)} = H^{(-)} - \frac{\cb}{2}, \qquad \ca^{\dagger} =
  \bh^{\dagger} \ah^{\dagger}, \qquad \ca = \ah \bh,  \label{eq:SSUSY} 
\end{equation}
where $h^{(1)}$ and $h^{(2)}$ are both partners of some intermediate Hamiltonian
\begin{equation*}
  h = \htilde^{(-)} + \frac{\cb}{2} = H^{(+)} - \frac{\cb}{2}
\end{equation*}
and the constant $\cb$ is related to the two factorization energies through
\begin{equation*}
  \cb = \et - E = \mp 2B.
\end{equation*}
On comparing Eq.\ (\ref{eq:intertwine}) with the two last relations in (\ref{eq:SSUSY}), we finally obtain
\begin{equation*}
  p(x) = \half \left(W + \wt\right) = - \frac{B\sinh x}{2B\cosh x - 2A - 1},
\end{equation*}
thereby completing the determination of the SSUSY transformation. It should be stressed here that since we have two possibilities for $\vt^{(-)}(x) = V^{(+)}(x)$ (see Eqs.\ (\ref{eq:V+bis}) and (\ref{eq:Vtilde})), there are two different paths for going from $V_{A,B}(x)$ to $V_{A,B,\rm{ext}}(x)$ or, in other words, the SSUSY transformation admits two different decompositions and therefore two different intermediate Hamiltonians $h$.\par
%
%=================================================================
%
\section{Generalizations}

The construction of extended potentials carried out in Sec.\ 2 in the GPT case can be generalized to some other exactly solvable potentials.\par
%
%++++++++++++++++++++++++++++++++++++++++++++++++++++++++++++++++++++
%
\subsection{Scarf I potential}

{}For the Scarf I potential
\begin{equation}
  V_{A,B}(x) = [A(A-1) + B^2] \sec^2 x - B (2A-1) \sec x \tan x, \quad - \frac{\pi}{2} < x < \frac{\pi}{2},
  \quad 0 < B < A-1,  \label{eq:SI-V}
\end{equation}
for instance, the bound-state wavefunctions being expressible in terms of Jacobi polynomials in the variable $z = \sin x$ \cite{cooper}, instead of (\ref{eq:W}) we would take $W(x) = a \tan x + b \sec x - \cos x (\sin x + c)^{-1}$ and obtain $c = - (2A-1)/(2B)$ for an extended potential without pole. As a consequence, we would get $V^{(+)}(x) = V_{A, B\pm1}(x)$ while its partner $V^{(-)}(x)$ would coincide with the extended Scarf I potential obtained in \cite{cq}. The isospectrality of the latter with $V_{A,B}(x)$, given in (\ref{eq:SI-V}), would then be explained by a SSUSY transformation entirely similar to that derived for the GPT potential.\par
%
%+++++++++++++++++++++++++++++++++++++++++++++++++++++++++++++++++
%
\subsection{\boldmath Scarf II potential and role of $\cal PT$ symmetry}

We may also inquire into what would happen for the corresponding hyperbolic Scarf II potential
\begin{equation}
  V_{A,B}(x) = [B^2 - A(A+1)] \sech^2 x + B (2A+1) \sech x \tanh x, \quad - \infty < x < \infty, \quad A>0, 
  \label{eq:SII-V}
\end{equation}
for which the corresponding variable $z$ is given by $z = \ri \sinh x$ \cite{cooper} and we would therefore assume $W(x) = a \tanh x + b \sech x - \ri \cosh x (\ri \sinh x + c)^{-1}$. Due to the fact that $\sinh x$ takes all real values whenever $x$ runs over the real line, neither of the two solutions for $c$, namely $c = - \ri (2A+1)/(2B)$ and $c = 2\ri B/(2A+1)$, would lead to an extended potential free from pole.\par
%
%-------------------------------------------------------------------------------------------------------------
%
It is interesting to note that this problem can be easily coped with by going to the $\cal PT$-symmetric Scarf II potential, obtained by replacing $B$ by $\ri B$ in (\ref{eq:SII-V}) \cite{bagchi00b, ahmed, levai, sinha, roy}. Depending on the value assumed for $c$, this would result in two distinct $\cal PT$-symmetric extended potentials, the first one being given by
\begin{equation*}
\begin{split}
  V_{A, \ri B, {\rm ext}}(x) &= - [B^2 + A(A+1)] \sech^2 x + \ri B (2A+1) \sech x \tanh x - \frac{2(2A+1)}
         {2A+1 - 2\ri B \sinh x}  \\
  & \quad + \frac{2[(2A+1)^2 - 4B^2]}{(2A+1 - 2\ri B \sinh x)^2} 
\end{split}
\end{equation*}
and the second one following from it  by permutation of $A+\half$ with $B$. To each of them, there would correspond a different SSUSY construct. Such a doubling of extensions hints at a connection with the existence of two series of real energy levels, $- (A-\nu)^2$, $\nu < A$, and $- \left(B-\half-\nu\right)^2$, $\nu < B-\half$, which characterizes the $\cal PT$-symmetric Scarf II potential whenever $A$ and $B-\half$ are both positive \cite{bagchi00b}.\par
%
%==================================================================
%
\section{Concluding remarks}

In conclusion, we have devised a SSUSY approach to isospectral conventional and rationally-extended exactly solvable potentials, which, at the same time as providing us with a constructive method for determining the latter, gives us a clue to calculating its polynomial solutions through the use of first-order differential operators. Such a procedure has been illustrated with the detailed example of the GPT potential and its application to the Scarf I potential of \cite{cq} has also been sketched. Furthermore, we have shown in the Scarf II case how $\cal PT$ symmetry may greatly help us in reconciling our approach with the requirement that the rationally-extended potential be singularity free in the domain where the conventional potential is defined.\par
%
%-------------------------------------------------------------------------------------------------------------------------
%
In this work, for simplicity's sake, we have restricted ourselves to some shape invariant potentials, whose eigenvalues are independent of one parameter, and we have chosen to change this independent parameter to reparametrize the potentials. It should be stressed that, as we plan to show elsewhere, our method is by no way limited to such cases.\par
%
%----------------------------------------------------------------------------------------------------------------------------
%
A further important point, which has not been dealt with here for lack of space but will also be considered in a future work, is the fact that the polynomials appearing in the rationally-extended potential bound-state wavefunctions  (see Eqs.\ (\ref{eq:definition}) and (\ref{eq:ext-psi})) belong to a class of exceptional orthogonal polynomials, which has been the topic of some recent mathematical study \cite{gomez08}. Our SSUSY approach provides us with a convenient way of constructing and generalizing such polynomials.
%
%====================================================================
%
\section*{Acknowledgments}

We are grateful to Dr.\ M.S.\ Plyushchay and Dr.\ T.\ Tanaka for useful correspondences. B.B.\ thanks Prof.\ M.P.\ Singh, Director, Ansal Institute of Technology, for his interest in this work. R.R.\ gratefully acknowledges the support of the National Fund for Scientific Research (FNRS), Belgium, and the warm hospitality at PNTPM, Universit\'e Libre de Bruxelles, where this work was carried out. He is also grateful to the Council of Scientific and Industrial Research (CSIR), New Delhi, for a grant (project no 21/0659/06/EMR-II).\par
%
%===================================================================
%

\end{document}